\documentstyle[12pt]{article}
\begin{document}

\noindent
{\large {\bf Strings and the Gauge Theory of Spacetime Defects}}\\ \\
\vskip0.5cm

\leftskip 1in
\noindent
{\bf Yishi Duan}\footnote{Institute of Theoretical Physics, Lanzhou
University, Lanzhou 730000, P. R. China}$^,$\footnote{E-mail:
ysduan@lzu.edu.cn} and {\bf Ying Jiang}$^{1,}$\footnote{Corresponding
author; E-mail: itp3@lzu.edu.cn}\\

\vskip0.5cm
\baselineskip22pt
\noindent
{\footnotesize
\hrulefill

\noindent
we present a new topological invariant to describe the space--time defect
which is closely related to torsion tensor in Riemann--Cartan manifold. By
virtue of the topological current theory and $\phi $--mapping method, we
show that there must exist many strings objects generated from the zero
points of $\phi $--mapping, and these strings are topological quantized and
the topological quantum numbers is the Winding numbers described by the Hopf
indices and the Brouwer degrees of the $\phi $--mapping.}\\
\_\hrulefill

\leftskip 0in
\vskip0.5cm
Key words: Topological quantization, string, torsion

\baselineskip22pt
\vskip1cm
\noindent
{\bf 1. INTRODUCTION}
\vskip0.5cm

The landscape of fundamental physics has changed substantially during the
last one or two decades (Rovelli, 1998). In fact, on the one hand at a
microscopical level the strong and weak interactions, while the
gravitational interaction is the weakest and seems not to play any role; on
the other hand, all known interactions, but gravitation, that is strong,
weak and electromagnetic interactions are well described within the
framework of relativistic quantum field theory in flat Minkowski space-time.
So at the first sight it seems that the gravitation has no effects when we
are concerned with elementary-particle physics. But today we know that is
not true (De Sabbata, 1994): in fact, if we consider the quantum theory in
curved instead of flat Minkowski spacetime we have some very important
effects (as, for instance, neutron interferometry (De Sabbata et.al 1991)),
and moreover when we go to a microphysical level, that is when we are
concerned with elementary-particle physics, we realize that the role of
gravitation becomes very important and necessary and this happens in the
first place when we consider the early universe or the Planck era. In this
unprecedented state of affairs, a large number of theoretical physicists
from different backgrounds have begun to address the piece of the puzzle
which is clearly missing: combining the two halves of the picture and
understanding the quantum properties of the gravitational field,
Equivalently, understanding the quantum properties of spacetime.

An exciting outcome of the interplay between particle physics and cosmology
is the string theory (Hindmarsh and Kibble, 1995). Strings are linear
defects, analogous to those topological defects found in some condensed
matter systems such as vortex lines in liquid helium, flux tubes in type-II
superconductors or disclination lines in liquid crystal, and it is closely
related to the torsion tensor of the Riemann--Cartan manifold (Duan et.al,
1994). String theory is strongly believed to solve the short-distance
problems of quantum gravity at the Plank scale by providing a fundamental
length $l_{str}=\sqrt{\hbar c/T}$, where $T$ is the string tension, and
provides a bridge between the physics of the very small and the very large.

In recent years, string theory has reached an exciting stage, where models
of various types (such as Wess-Zumino-Witten model (Bakas, 1993, 1994),
Ramond-Ramond charges of type II string theory (Mirjam Cvetic and Donam
Youm, 1996) and supersymmetric $SO(10)$ model (Jeannerot, 1996)) have been
of much interest in differential geometry (Bakas and Stetsos, 1996), field
theory (Robinson and Ziabicki, 1996), and general relativity (Larsen and
S\'anchez, 1996). Though all these features make string theory very
attractive, but since most of them are based on some concrete models, they
are not very perfect and the topological property of strings are not very
clear yet.

As is well known, torsion is slight modification of the Einstein theory of
relativity (proposed in the 1922-23 by E. Cartan (1922)), but is
generalization that appears to be necessary when one tries to conciliate
general relativity with quantum theory. The main purpose of this paper is to
establish a topological theory for string through $\phi $--mapping method
(Duan and Meng, 1993) and the theory of composed gauge potential (Duan and
Lee, 1995) in 4-dimensional Riemann--Cartan manifold $X$ without any
concrete models in early universe. This theoretical framework includes three
basic aspects: the generation of multistring in a 4--dimensional
Riemann--Cartan manifold, the topological charges of the multistrings and
their evolution equations. 

This paper is organized as follows: In section 2, we introduce a new
topological invariant to describe the spacetime defect, and using the
decomposition of gauge potential, we get the inner structure of torsion. In
section 3, by means of the topological tensor current theory and the $\phi $%
--mapping method, the multistrings are generated naturally at the zero
points of the vector total field $\vec \phi $ and the topological quantum
numbers of the length of these strings are just the Hopf indices and the
Brouwer degrees of $\phi $-mapping. Furthermore, using some important
relations, the Lagrangian density of many strings is obtained and the
corresponding evolution equations are deduced in section 4, and it is
pointed out that the Lagrangian density is a generalization of that of
Nielsen for strings and the evolution equations relate to the Hamornic
mapping in general relativity.

\vskip1cm

\noindent
{\bf 2. NEW TOPOLOGICAL INVARIANT AND SPACETIME DEFECT}
\vskip0.5cm

Using vierbein theory and the gauge potential decomposition, we will
construct the Invariant formulation of the spacetime defects. The defects of
space-time has been discussed from different points of view by many
physicists. We will follow Duan and Zhang (1990) in which the defect of
spacetime was studied from the point of view of gauge field theory. The
dislocation is described by the torsion 
\begin{equation}
\label{t}T_{\mu \nu }^A=D_\mu e_\nu ^A-D_\nu e_\mu ^A,\;\;\;\mu ,\nu
,A=1,2,3,4 
\end{equation}
where $e_\mu ^A$ is the vierbein field and its gauge covariant derivative 
$$
D_\mu e_v^A=\partial _\mu e_v^A-\omega _\mu ^{AB}(x)e_v^B, 
$$
where $\omega _\mu ^{AB}$ stands for the spin connection of the Lorentz
group.

By analogy with the 't Hooft's viewpoint ('tHooft, 1974), to establish a
physical observable theory of space--time defect, we must first define a
gauge invariant antisymmetrical 2--order tensor from torsion tensor with
respect of a unit vector field $N^A(x)$ as follows%
$$
T_{\mu \nu }=T_{\mu \nu }^AN^A+e_\nu ^AD_\mu N^A-e_\mu ^AD_\nu N^A. 
$$
By making use of 
$$
D_\mu N^A=\partial _\mu N^A-\omega _\mu ^{AB}N^B 
$$
and (\ref{t}), it can be rewritten as 
\begin{equation}
T_{\mu \nu }=\partial _\mu A_\nu -\partial _\nu A_\mu 
\end{equation}
where $A_\mu =e_\mu ^AN^A$ is a kind of $U(1)$ gauge potential. This shows
that the antisymmetrical tensor $T_{\mu \nu }$ expressed in terms of $A_\mu $
is just the $U(1)$ like gauge field strength (i.e. the curvature on $U(1)$
principle bundle with base manifold $X$ ), which is invariant for the $U(1)$%
--like gauge transformation 
\begin{equation}
\label{cc}A_\mu ^{\prime }(x)=A_\mu (x)+\partial _\mu \Lambda (x)
\end{equation}
where $\Lambda (x)$ is an arbitrary function.

Now, let us investigate the integral of the two--form $T=\frac 12T_{\mu \nu
}dx^\mu \wedge dx^\nu $, which will be shown that, in topology, it is
associated with the first Chern class (but not!), i.e.%
$$
l=\int \frac 12T_{\mu \nu }dx^\mu \wedge dx^\nu . 
$$
Since the integral quantity $l$ carries neither the coordinate index nor the
group index, it must be pointed that $l$ is invariant under general
coordinate transformation as well as local Lorentz transformation.
Furthermore, in $l$ there is another $U(1)$--like gauge invariance for (\ref
{cc}). In fact, $l$ is a new topological invariant and is used to measure
the size of defects of the spacetime with the dimension of length.

Very commonly, topological property of a physical system is much more
important and worth investigating mediculously. It is our conviction that,
in order to get a topological result, one should input the topological
information from the beginning. Two useful tools---$\phi $-mapping method
and composed gauge potential theory--- just do the work. As $A_\mu $ is a
kind of $U(1)$ gauge potential, for a section $\Phi (x)$ of the complex line
bundle $L(X)$ with the base manifold $X$ which is looked upon as the ordered
parameter of the spacetime defects, the corresponding $U(1)$ covariant
derivative of $\Phi (x)$ with $A_\mu $ is%
$$
D_\mu \Phi (x)=\partial _\mu \Phi (x)-i\frac{2\pi }{L_p}A_\mu \Phi (x), 
$$
$$
D_\mu \Phi ^{*}(x)=\partial _\mu \Phi ^{*}(x)+i\frac{2\pi }{L_p}A_\mu \Phi
^{*}(x), 
$$
where $L_p$ is the Plank length introduced to make the both sides of the
formula with the same dimension (Duan et.al, 1994). From the above
equations, $A_\mu (x)$ can be expressed by 
\begin{equation}
A_\mu (x)=\frac{iL_p}{4\pi \Phi \Phi ^{*}}[(\Phi \partial _\mu \Phi
^{*}-\Phi ^{*}\partial _\mu \Phi )-(\Phi D_\mu \Phi ^{*}-\Phi ^{*}D_\mu \Phi
)].
\end{equation}
More deeper calculation can draw the result that 
\begin{equation}
\label{1}A_\mu (x)=\frac{iL_p}{4\pi }(\frac \Phi {\sqrt{\Phi \Phi ^{*}}%
}\partial _\mu \frac{\Phi ^{*}}{\sqrt{\Phi \Phi ^{*}}}-\frac{\Phi ^{*}}{%
\sqrt{\Phi \Phi ^{*}}}\partial _\mu \frac \Phi {\sqrt{\Phi \Phi ^{*}}})-%
\frac{iL_p}{4\pi \Phi \Phi ^{*}}(\Phi D_\mu \Phi ^{*}-\Phi ^{*}D_\mu \Phi ).
\end{equation}
From the well--known Chern--Weil homomorphism (Nash and Sen, 1983), we know
that our new topological invariant is independent of the gauge potential,
i.e. the last term in RHS of the equation (\ref{1}) has nothing to do with
the topological property in our theory. So we have many choice of gauge
potential and the choice depends on the convenience of calculus. In the
present work, we select $A_\mu $ as%
$$
A_\mu (x)=\frac{iL_p}{4\pi }(\frac \Phi {\sqrt{\Phi \Phi ^{*}}}\partial _\mu 
\frac{\Phi ^{*}}{\sqrt{\Phi \Phi ^{*}}}-\frac{\Phi ^{*}}{\sqrt{\Phi \Phi ^{*}%
}}\partial _\mu \frac \Phi {\sqrt{\Phi \Phi ^{*}}}) 
$$
satisfying the relation (\ref{cc}) for the corresponding $U(1)$ gauge
transformation $\Phi ^{\prime }(x)=\Lambda (x)\Phi (x)$. As is well known,
the section $\Phi (x)$ of the complex line bundle can be expressed by%
$$
\Phi (x)=\phi ^1(x)+i\phi ^2(x), 
$$
i.e. the section of the complex line bundle is equivalent to a 2-dimensional
real vector field $\vec \phi =(\phi ^1,\phi ^2)$, and $\sqrt{\Phi \Phi ^{*}}%
=||\phi ||=\sqrt{\phi ^a\phi ^a}(a=1,2)$. By defining the direction of the
vector field $\vec \phi $ as 
\begin{equation}
\label{n}n^a(x)=\frac{\phi ^a(x)}{||\phi (x)||},
\end{equation}
we can obtain the expression of $A_\mu (x)$ in terms of $n^a$ from (\ref{1}) 
\begin{equation}
\label{star}A_\mu (x)=\frac{L_p}{2\pi }\epsilon _{ab}n^a(x)\partial _\mu
n^b(x).
\end{equation}
Obviously, $n^a(x)n^a(x)=1,$ and $n^a(x)$ is a section of the sphere bundle $%
S\left( X\right) $ (Duan and Meng, 1993). The zero points of $\phi ^a(x)$
are just the singular points of $n^a(x).$ Thus we get the total
decomposition of $U(1)$ gauge potential $A_\mu $ with the unit 2-vector
field $n^a$, and because of the topological property of $n^a$, we input the
topological information successfully.

\vskip1cm
\noindent
{\bf 3. 2-ORDER TOPOLOGICAL TENSOR CURRENT AND THE GENERATION OF STRINGS ON
RIEMANN--CARTAN MANIFOLD}

\vskip0.5cm

In recent years, the topological current theory proposed by Duan(one of the
present authors) has been found to play a significant role in particle
physics, field theory(especially the gauge theory) (Duan and Meng, 1993;
Duan and Lee, 1995; Duan and Zhang, 1990), and the topological current
theory can only be used to discussed the motion of point-like particles( or
point-like singularity). In order to study the string theory, we should
extend the concept, and introduce a topological tensor current of second
order from torsion.

From the above discussions, we can define a dual tensor $j^{\mu \nu }$ of $%
T_{\lambda \rho }$ as follow%
$$
j^{\mu \nu }=\frac 12\frac 1{\sqrt{g_x}}\epsilon ^{\mu \nu \lambda \rho
}T_{\lambda \rho } 
$$
\begin{equation}
\label{12}=\frac 12\frac 1{\sqrt{g_x}}\epsilon ^{\mu \nu \lambda \rho
}(\partial _\lambda A_\rho -\partial _\rho A_\lambda ). 
\end{equation}
With the decomposition of $A_\mu $ in (\ref{star}), $j^{\mu \nu }$ can be
expressed in terms of $n^a$ by 
\begin{equation}
\label{13}j^{\mu \nu }=\frac{L_p}{2\pi }\frac 1{\sqrt{g_x}}\epsilon ^{\mu
\nu \lambda \rho }\epsilon _{ab}\partial _\lambda n^a\partial _\rho n^b, 
\end{equation}
which shows that $j^{\mu \nu }$ is just a 2-order topological tensor current
satisfying%
$$
j^{\mu \nu }=-j^{\nu \mu },\;\;\;\;\;\frac 1{\sqrt{g_x}}\partial _\mu (\sqrt{%
g_x}j^{\mu \nu })=0, 
$$
i.e. $j^{\mu \nu }$ is antisymmetric and identically conserved.

Using (\ref{13}) and 
$$
\partial _\mu n^a=\frac 1{\Vert \phi \Vert }\partial _\mu \phi ^a+\phi
^a\partial _\mu (\frac 1{\Vert \phi \Vert }),\;\;\;\;\;\;\;\frac \partial
{\partial \phi ^a}(\ln \Vert \phi \Vert )=\frac{\phi ^a}{\Vert \phi \Vert ^2}%
, 
$$
which should be looked upon as generalized functions, $j^{\mu \nu }$ can be
expressed by 
\begin{equation}
\label{14}j^{\mu \nu }=\frac{L_p}{2\pi }\frac 1{\sqrt{g_x}}\epsilon ^{\mu
\nu \lambda \rho }\epsilon _{ab}\frac \partial {\partial \phi ^c}\frac
\partial {\partial \phi ^a}(\ln \Vert \phi \Vert )\partial _\lambda \phi
^c\partial _\rho \phi ^b. 
\end{equation}
By defining the general Jacobian determinants $J^{\mu \nu }(\frac \phi x)$
as 
\begin{equation}
\label{15}\epsilon ^{ab}J^{\mu \nu }(\frac \phi x)=\epsilon ^{\mu \nu
\lambda \rho }\partial _\lambda \phi ^a\partial _\rho \phi ^b 
\end{equation}
and making use of the Laplacian relation in $\phi $-space%
$$
\partial _a\partial _a\ln \Vert \phi \Vert =2\pi \delta (\vec \phi
),\;\;\;\;\;\;\;\;\partial _a=\frac \partial {\partial \phi ^a}, 
$$
we obtain the $\delta $-like topological tensor current rigorously 
\begin{equation}
\label{16}j^{\mu \nu }=\frac 1{\sqrt{g_x}}L_p\delta (\vec \phi )J^{\mu \nu
}(\frac \phi x). 
\end{equation}
It is obvious that $j^{\mu \nu }$ is non-zero only when $\vec \phi =0.$

Suppose that for the system of equations 
$$
\phi ^1(x)=0,\,\,\,\,\phi ^2(x)=0, 
$$
there are $l$ different solutions, when the solutions are regular solutions
of $\phi $ at which the rank of the Jacobian matrix $[\partial _\mu \phi ^a]$
is 2, the solutions of $\vec \phi (x)=0$ can be expressed parameterizedly by 
\begin{equation}
\label{5}x^\mu =z_i^\mu (u^1,u^2),\,\,\,i=1,\cdot \cdot \cdot l, 
\end{equation}
where the subscript $i$ represents the $i$-th solution and the parameters $%
u^I(I=1,2)$ span a 2-dimensional submanifold with the metric tensor $%
g_{IJ}=g_{\mu \nu }\frac{\partial x^\mu }{\partial u^I}\frac{\partial x^\nu 
}{\partial u^J}$which is called the $i$-th singular submanifold $N_i$ in $X.$
For each $N_i$, we can define a normal submanifold $M_i$ in $X$ which is
spanned by the parameters $v^A(A=1,2)$ with the metric tensor $g_{AB}=g_{\mu
\nu }\frac{\partial x^\mu }{\partial v^A}\frac{\partial x^\nu }{\partial v^B}
$, and the intersection point of $M_i$ and $N_i$ is denoted by $p_i$. By
virtue of the implicit function theorem, at the regular point $p_i$, it
should be hold true that the Jacobian matrix $J(\frac \phi v)$ satisfies 
\begin{equation}
J(\frac \phi v)=\frac{D(\phi ^1,\phi ^2)}{D(v^1,v^2)}\neq 0. 
\end{equation}

As is well known (Schouten, 1951), the definition of the $\delta $ function
on a submanifold $N_i$ $\delta (N_i)$ should be satisfied the surface area
relation%
$$
\int \delta (N_i)\sqrt{g_x}d^4x=\int_{N_i}\sqrt{g_u}d^2u, 
$$
where $\sqrt{g_x}d^4x$ and $\sqrt{g_u}d^2u$ ($\sqrt{g_u}=\det (g_{IJ})$)are
invariant volume element of $X$ and $N_i$ respectively, and the expression
of $\delta (N_i)$ is%
$$
\delta (N_i)=\int_{N_i}\frac 1{\sqrt{g_x}}\delta ^4(\vec x-\vec z_i(u^1,u^2))%
\sqrt{g_u}d^2u. 
$$
Following this, by analogy with the procedure of deducing $\delta (f(x))$,
since 
\begin{equation}
\delta (\vec \phi )=\left\{ 
\begin{array}{cc}
+\infty , & for\;\vec \phi (x)=0 \\ 
0, & for\;\vec \phi (x)\neq 0
\end{array}
\right. =\left\{ 
\begin{array}{cc}
+\infty , & for\;x\in N_i \\ 
0, & for\;x\notin N_i
\end{array}
\right. ,
\end{equation}
we can expand the $\delta $--function $\delta (\vec \phi )$ as 
\begin{equation}
\label{delta}\delta (\vec \phi )=\sum_{i=1}^lc_i\delta (N_i),
\end{equation}
where the coefficients $c_i$ must be positive, i.e. $c_i=\mid c_i\mid $.

In the following, we will discuss the dynamic form of the tensor current $%
j^{\mu \nu }$ and study the topological quantization of strings through the
Winding numbers (Guillemin and Pollack, 1974) $W_i$ of $\vec \phi $ on $M_i$
at $p_i$%
$$
W_i=\frac 1{2\pi }\int_{\partial \Sigma _i}d\arctan [\frac{\phi ^2}{\phi ^1}%
], 
$$
where $\partial \Sigma _i$ is the boundary of a neighborhood $\Sigma _i$ of $%
p_i$ on $M_i$with $p_i\notin \partial \Sigma _i$. It is well-known that the
Winding numbers $W_i$ are corresponding to the first homotopy group $\pi
[S^1]=Z$ (the set of integers). By making use of ($\ref{n}$), it can be
precisely proved that 
\begin{equation}
\label{wi}W_i=\frac 1{2\pi }\int_{\partial \Sigma _i}n^{*}(\epsilon
_{ab}n^adn^b),
\end{equation}
where $n^{*}$ is the pull back of map $n$. This is another definition of $W_i
$ by the Gauss map (Dubrosin et.al, 1985) $n:\partial \Sigma _i\rightarrow
S^1.$ In topology it means that, when the point $v=(v^1,v^2)$ covers $%
\partial \Sigma _i$ once, the unit vector $n^a$ will cover $S^1$ $W_i$
times, which is a topological invariant and is also called the degree of
Gauss map. Using the Stokes' theorem in the exterior differential form and (%
\ref{wi}), one can deduce that%
$$
W_i=\frac 1{2\pi }\int_{\Sigma _i}\epsilon _{ab}\partial _An^a\partial
_Bn^bdv^A\wedge dv^B 
$$
$$
=\frac 1{2\pi }\int_{\Sigma _i}\epsilon ^{AB}\epsilon _{ab}\partial
_An^a\partial _Bn^bd^2v. 
$$
Then, by duplicating the above process, we have 
\begin{equation}
\label{W}W_i=\int_{\Sigma _i}\delta (\vec \phi )J(\frac \phi v)d^2v,
\end{equation}
substituting (\ref{delta}) into (\ref{W}), and considering that only one $%
p_i\in \Sigma _i$, we can get%
$$
W_i=\int_{\Sigma _i}c_i\delta (N_i)J(\frac \phi v)d^2v 
$$
$$
=\int_{\Sigma _i}\int_{N_i}c_i\frac 1{\sqrt{g_x}\sqrt{g_v}}\delta ^4(\vec
x-\vec z_i(u^1,u^2))J(\frac \phi v)\sqrt{g_u}d^2u\sqrt{g_v}d^2v, 
$$
where $\sqrt{g_v}=\det (g_{AB})$. Because $\sqrt{g_u}\sqrt{g_v}d^2ud^2v$ is
the invariant volume element of the Product manifold $M_i\times N_i$, so it
can be rewritten as $\sqrt{g_x}d^4x$. Thus, by calculating the integral and
with positivity of $c_i$, we get 
\begin{equation}
c_i=\frac{\beta _i\sqrt{g_v}}{\mid J(\frac \phi v)_{p_i}\mid }=\frac{\beta
_i\eta _i\sqrt{g_v}}{J(\frac \phi v)_{p_i}},
\end{equation}
where $\beta _i=|W_i|$ is a positive integer called the Hopf index (Milnor,
1965) of $\phi $-mapping on $M_i,$ it means that when the point $v$ covers
the neighborhood of the zero point $p_i$ once, the function $\vec \phi $
covers the corresponding region in $\vec \phi $-space $\beta _i$ times, and $%
\eta _i=signJ(\frac \phi v)_{p_i}=\pm 1$ is the Brouwer degree (Milnor,
1965) of $\phi $-mapping. Substituting this expression of $c_i$ and (\ref
{delta}) in (\ref{16}), we gain the total expansion of the string current%
$$
j^{\mu \nu }=\frac{L_p}{\sqrt{g_x}}\sum_{i=1}^l\frac{\beta _i\eta _i\sqrt{g_v%
}}{J(\frac \phi v)|_{p_i}}\delta (N_i)J^{\mu \nu }(\frac \phi x) 
$$
From the above equation, we conclude that the inner structure of $j^{\mu \nu
}$ is labelled by the total expansion of $\delta (\vec \phi )$, which
includes the topological information $\beta _i$ and $\eta _i.$

With the discovery of an explicit four-particle amplitude that combines the
narrow-resonance approximation with Regge behavior and crossing symmetry,
some physicists began to study the dual resonance models, i.e. string model,
which can be generated from our topological tensor current theory. It is
obvious that, in (\ref{5}), when $u^1$ and$\,u^2$ are taken to be time-like
evolution parameter and space-like string parameter, respectively, the inner
structure of $j^{\mu \nu }$ just represents $l$ strings moving in the
4--dimensional Riemann-Cartan manifold $X$. The 2-dimensional singular
submanifolds $N_i\,\,(i=1,\cdot \cdot \cdot l)$ are their world sheets. Here
we see that the strings are generated from where $\vec \phi =0$ and does not
tie on any concrete models. Furthermore, we see that the Hopf indices $\beta
_i$ and Brouwer degree $\eta _i$ classify these strings. In detail, the Hopf
indices $\beta _i$ characterize the absolute values of the topological
quantization and the Brouwer degrees $\eta _i=+1$ correspond to strings
while $\eta _i=-1$ to antistrings.

\vskip1cm
\noindent
{\bf 4. THE EVOLUTION EQUATIONS OF STRINGS}

\vskip0.5cm

At the beginning of this section, we firstly give some useful relations to
study many strings theory. On the $i$-th singular submanifold $N_i$ we have%
$$
\phi ^a(x)|_{N_i}=\phi ^a(z_i^1(u),\cdot \cdot \cdot z_i^4(u))\equiv 0, 
$$
which leads to%
$$
\partial _\mu \phi ^a\frac{\partial x^\mu }{\partial u^I}|_{N_i}=0,\;\,\;\;%
\;\;I=1,2. 
$$
Using this relation and the expression of the Jacobian matrix $J(\frac \phi
v)$, we can obtain 
$$
J^{\mu \nu }(\frac \phi x)|_{\vec \phi =0}=\frac 12\epsilon ^{\mu \nu
\lambda \rho }\epsilon _{ab}\frac{\partial \phi ^a}{\partial x^\lambda }%
\frac{\partial \phi ^b}{\partial x^\rho } 
$$
$$
=\frac 12\epsilon ^{\mu \nu \lambda \rho }\epsilon _{ab}\frac{\partial \phi
^a}{\partial v^A}\frac{\partial \phi ^b}{\partial v^B}\frac{\partial v^A}{%
\partial x^\lambda }\frac{\partial v^B}{\partial x^\rho } 
$$
\begin{equation}
\label{35}=\frac 12\epsilon ^{\mu \nu \lambda \rho }\epsilon _{AB}J(\frac
\phi v)\frac{\partial v^A}{\partial x^\lambda }\frac{\partial v^B}{\partial
x^\rho }, 
\end{equation}
then from this expression, the rank--two tensor current can be expressed by 
\begin{equation}
\label{36}j^{\mu \nu }=\frac{L_p}{2\sqrt{g_x}}\sum_{i=1}^l\beta _i\eta _i%
\sqrt{g_v}\delta (N_i)\epsilon ^{\mu \nu \lambda \rho }\epsilon _{AB}\frac{%
\partial v^A}{\partial x^\lambda }\frac{\partial v^B}{\partial x^\rho }. 
\end{equation}

Corresponding to the rank--two topological tensor currents $j^{\mu \nu }$,
it is easy to see that the Lagrangian of many strings is just%
$$
L=\frac 1{L_p}\sqrt{\frac 12g_{\mu \lambda }g_{\nu \rho }j^{\mu \nu
}j^{\lambda \rho }}=\delta (\vec \phi ) 
$$
which includes the total information of strings in $X$ and is the
generalization of Nielsen's Lagrangian for string (Nielsen and Olesen,
1973). The action in $X$ is expressed by 
$$
S=\int_XL\sqrt{g_x}d^4x=\sum_{i=1}^l\beta _i\eta _i\int_{N_i}\sqrt{g_u}%
d^2u=\sum_{i=1}^l\beta _i\eta _iS_i 
$$
where $S_i$ is the area of the singular manifold $N_i$. It must be
pointed out here that the Nambu--Goto
action (Nambu, 1970; Forster, 1974; Orland, 1994), which is the basis of
many works on string theory, is derived naturally from our theory. From the
principle of least action, we obtain the evolution equations of many strings 
\begin{equation}
\label{38}g^{IJ}\frac{\partial g_{\nu \lambda }}{\partial x^\mu }\frac{%
\partial x^\nu }{\partial u^I}\frac{\partial x^\lambda }{\partial u^J}%
-2\frac 1{\sqrt{g_u}}\frac \partial {\partial u^I}(\sqrt{g_u}g^{IJ}g_{\mu
\nu }\frac{\partial x^\nu }{\partial u^J})=0,\;\;\;I,J=1,2.
\end{equation}
As a matter of fact, this is just the equation of harmonic map (Duan et.al,
1992).

\vskip1cm
\noindent
{\bf 5. CONCLUSION}

\vskip0.5cm

In summary, we have studied the topological quantization of the strings in
Riemann--Cartan space--time by making use of the composed gauge theory and
the $\phi$--mapping topological current theory. As a result, the strings are
generated from the zero points of $\phi$--mapping and the topological
quantum numbers of these strings are the Winding numbers which are
determined by the Hopf indices and the Brouwer degrees of $\phi$--mapping,
the singular manifolds of $\vec \phi$ are just the evolution surfaces of
these strings. The whole theory in this paper is not only covariant under
general coordinate transformations but also gauge invariant.

\vskip1cm
\noindent
{\bf ACKNOWLEDGMENT}

\vskip0.5cm
This work is supported by the National Natural Science
Foundation of P. R. China.

\vskip1cm
\noindent
{\bf REFERENCES}

\vskip0.5cm
{\footnotesize \baselineskip22pt \noindent
Bakas, I. \ {\it Phys. Lett.} {\bf B319}, 457 (1993); \ {\it Int. J. Mod.
Phys.} {\bf A9}, 3443 (1994) }

{\footnotesize \noindent
Bakas, I. and Sfetsos, K. \ {\it Phys. Rev. } {\bf D54}, 3995 (1996) }

{\footnotesize \noindent
Cartan, \`E. \ {\it C. R. Acad. Sci. {\bf 174}, }593 (1922) }

{\footnotesize \noindent
De Sabbata, V. \ {\it IL Nuovo Cimento }{\bf 107A}, 363 (1994) }

{\footnotesize \noindent
De Sabbata, V., et. al. \ {\it Int. J. Theor. Phys}. {\bf 30}, 167 (1991) }

{\footnotesize \noindent
Duan, Y. S. et al\ {\it Gen. Rel. Grav}. {\bf 24}, 1033 (1992) }

{\footnotesize \noindent
Duan, Y. S. et al \ {\it J. Math. Phys. }{\bf 35}, 1 (1994) }

{\footnotesize \noindent
Duan, Y. S. and Lee, X. G. \ {\it Helv. Phys. Acta}{\bf \ 68}, 513 (1995) }

{\footnotesize \noindent
Duan, Y. S. and Meng, X. H. \ {\it J. Math. Phys. }{\bf 34}, 1549 (1993) }

{\footnotesize \noindent
Duan, Y. S. and Zhang, S. L. \ {\it Int. J. Eng. Sci}. {\bf 28}, 689 (1990) }

{\footnotesize \noindent
Dubrosin, B. A. et. al., \ {\it Modern Geometry Method and Application},
Part II, Springer-Verlag }

{\footnotesize New York Inc. 1985 }

{\footnotesize \noindent
Forster, D. {\it \ Nucl. Phys. {\bf B81}, }84 (1974) }

{\footnotesize \noindent
Guillemin, V. and Pollack, A. \ {\it Differential Topology}, (Prentice-Hall,
Inc. Englewood Cliffs, }

{\footnotesize New Jersey, 1974) }

{\footnotesize \noindent
Hindmarsh, M. B. and Kibble, T. W. B. \ {\it Rep. Prog. Phys. }{\bf 58}, 477
(1995) }

{\footnotesize \noindent
Jeannerot, R. \ {\it Phys. Rev.} {\bf D53}, 5426 (1996). }

{\footnotesize \noindent
Larsen, A. L. and S\'anchez, N. \ {\it Phys. Rev.} {\bf D54}, 2483 (1996); 
{\bf D54}, 2801 (1996). }

{\footnotesize \noindent
Milnor, J. W. \ Topology, From the Differential Viewpoint, The Unviersity
Press }

{\footnotesize of Virginia Charlottesville, 1965 }

{\footnotesize \noindent
Mirjam Cveti\v c and Donam Youm,\ {\it Phys. Rev.} {\bf D54}, 2612 (1996). }

{\footnotesize \noindent
Nambu, Y. \ {\it Lectures at the Copenhagen, Symposium }(1970) }

{\footnotesize \noindent
Nash, S. and Sen, S. \ {\it Topology and Geometry for Physicists.} Academic
Pre. Inc. London (1983) }

{\footnotesize \noindent
Nielsen, H. B. and Olesen, P. \ {\it Nucl. Phys}. {\bf B57}, 367 (1973) }

{\footnotesize \noindent
Orland, P. {\it \ Nucl. Phys. {\bf B428}, }221 (1994) }

{\footnotesize \noindent
Robinson, M. and Ziabicki, J. \ {\it Phys. Rev.} {\bf D53}, 5924 (1996) }

{\footnotesize \noindent
Rovelli, C. \ {\it Strings, Loops and others: a criticla survey of the
present approaches to quantum gravity} }

{\footnotesize (gr-qc/9803024) }

{\footnotesize \noindent
Schouten, J. A. {\it Tensor Analysis for Physicists} (Oxford, Clarendon
Press, 1951) }

{\footnotesize \noindent
'tHooft, G. \ {\it Nucl. Phys.} {\bf b79}, 276 (1974) }

\end{document}